\documentclass{mem}
\usepackage{natbib}\usepackage{txfonts}\usepackage{balance}
\usepackage{graphicx}
\usepackage[a4paper,breaklinks]{hyperref}
\idline{75}{282}
\begin{document}
\def\teff{$T\rm_{eff }$}
\def\kms{$\mathrm {km s}^{-1}$}

\title{
Kinematics and Luminosities of Brown Dwarfs with the BDNYC group
}

   \subtitle{}

\author{
A. Riedel\inst{1,2} 
          }

\institute{
Physics Department, Hunter College
695 Park Avenue, New York, New York 10065
\and
The American Museum of Natural History
200 Central Park West, New York, New York 10024
\email{adric.riedel@gmail.com}
}

\authorrunning{Riedel}

\titlerunning{Kinematics of Brown Dwarfs}

\abstract{
Due to magnitude limits, the Gaia survey will not delve as deeply into the local population of brown dwarfs as it will other stellar populations.  While hundreds or thousands of brown dwarfs will be measured by Gaia, we propose a different, indirect method wherein studies using Gaia data will help teach us about brown dwarfs: Identifying moving groups that contain brown dwarfs.  This use of Gaia data will directly help attempts to disentangle the effects of age and mass on brown dwarf spectra, which opens the possibilities for determining empirical constraints on brown dwarf evolution.
}
\maketitle{}

\section{Introduction}

Brown dwarfs are by def\/inition starlike objects that never attain core temperatures suff\/icient to sustain fusion of hydrogen into helium due to their low masses.  Because they do not sustain hydrogen fusion, they are not stars.  Because of the continuum of masses and temperatures that result from stellar formation processes, brown dwarfs nevertheless resemble stars in many ways, and form a bridge between stars and gas giant planets.  Without a fusion power source, they continuously shrink and cool, rather than falling onto a main sequence.  Thus, we are left with a situation where it is diff\/icult to tell if a brown dwarf of a given temperature is a young, low-mass brown dwarf, or an old, high-mass brown dwarf.  The discrepancy in apparent gravity and age makes it diff\/icult to make evolutionary tracks without assumptions about the internal physics of the brown dwarfs.

To solve this problem, we must break the age-mass degeneracy.  This can be done by measuring the brown dwarf masses -- preferably dynamical masses, although brown dwarfs in binaries are notoriously rare -- or it can be done by measuring the ages of the brown dwarfs.  This can be done by connecting them to a group of stars with a known age.  The nearby young moving groups are expected to be coeval groups of a few hundred stars, the products of single small bursts of star formation in the process of dissipating into the galactic disk.  They are young enough to exhibit a wide range of brown dwarf evolutionary states, and are close enough that particularly low-mass objects are bright enough for serious study.

\section{Moving Group Kinematics}

Historically, f\/inding young stars was a tricky business.  The closest star-forming regions (i.e. Taurus-Auriga, Scorpius-Centaurus, Orion) are not very near, and the nearest groups (The Hyades, Ursa Major) were not particularly young.  In recent decades, ``Isolated T Tauri stars'' have been found, and there are now at least 20 proposed groupings of young (less than 150 Myr old) stars near the Sun, many of which are now believed to contain brown dwarfs.  At present, the ages memberships are not settled, though many authors (most recently \citet{Malo2013} and \citet{Gagne2014}) have attempted to derive lists of true consistent members.

\subsection{Moving Group Kinematics Codes}
The workhorse technique for identifying memberships is kinematics.  Other methods -- spectroscopic, astrometric, and photometric -- can provide precise estimates of the ages of systems, but only kinematics can identify a particular moving group.

The method analyzed here is a semi-convergence method.  It considers up to three metrics (proper motion, parallax, and radial velocity as available; spatial positions are not considered) to determine memberships. The methodology is much like a classical convergence code (as in \citealt{Rodriguez2013}) in that it operates by comparing a predicted proper motion vector (for a given moving group, at the RA and DEC of the target star) to the measured proper motion vector of the target object.  The difference is that the semi-convergence method uses the UVW matricies from \citet{Johnson1987} to convert UVW space velocities to $\mu_{RA,pred}$, $\mu_{DEC,pred}$ and $RV_{pred}$ rather than calculating the vector from the convergent point of the group. 

From that point onward, the methods are similar:  in both cases, the observed motion of the system $\mu_{RA}$, $\mu_{DEC}$ is split up into $\mu_{parallel}$ and $\mu_{perpendicular}$ components for analysis, where the perpendicular component should be zero if the system is a perfect match to the group.  The magnitude of the proper motion vector can be used to derive a kinematic distance; in the same way, the expected $RV_{pred}$ can be compared to a measured $RV$ for a third goodness-of-f\/it estimate.

\begin{figure}[]
\resizebox{\hsize}{!}{\includegraphics[clip=true, width=0.5\textwidth]{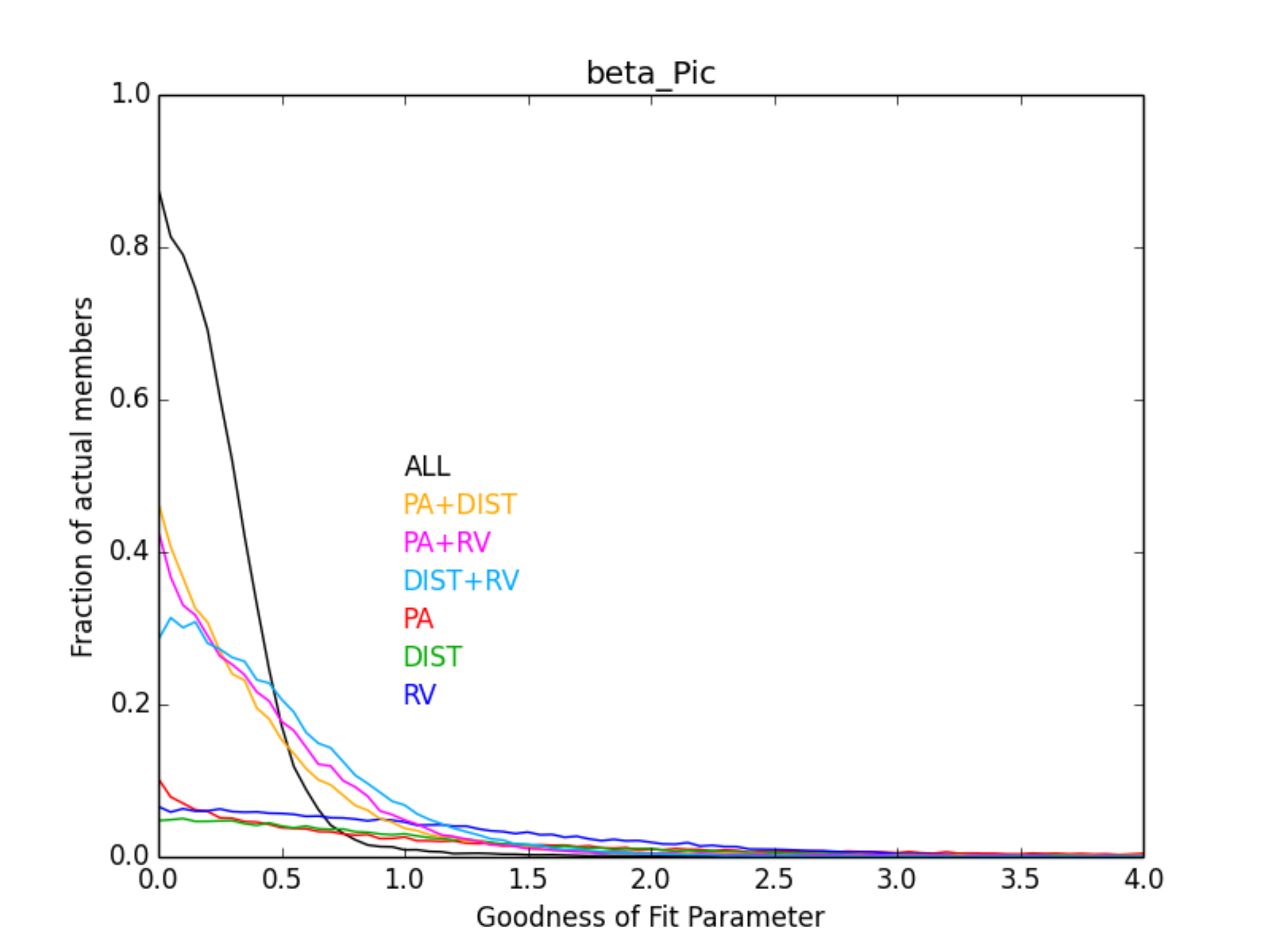}}
\caption{
\footnotesize
Plot of the fraction of members of $\beta$ Pictoris as a function of the goodness-of-f\/it parameter, for all combinations of available data.
}
\label{fig:betapic_all}
\end{figure}

Converting the magnitude of the perpendicular component of the proper motion vector to a probability of membership required calibration. This calibration stage was made by drawing f\/ive million points from the 6-dimensional distributions of the various known nearby moving groups, populated according to the size of each group.  As an example, f\/ield stars are roughly 15 times more common (137 young stars among 2167 star systems within 25 pc, Henry et al. in prep) than young stars, and they accounted for 15 times as many draws as all young stars put together.  The probabilities of membership were determined by the fraction of draws that were ``actual'' members, as a function of the combined goodness-of-f\/it parameter (Figure \ref{fig:betapic_all}).

\subsection{Lessons Learned from Moving Group Kinematics codes}

In many cases it is impossible to be 100\% certain that an object is a member by kinematics alone. Figure \ref{fig:betapic_all} shows the ability of the code to identify members of $\beta$ Pictoris. If we have only the proper motion (red line) as a parameter, even a perfect match to $\beta$ Pictoris has only a 10\% chance of actually being a member of $\beta$ Pictoris.  Adding in a trigonometric parallax or radial velocity boosts the maximum certainty to just over 40\%, but even when using the $\mu$, $\pi$, and $RV$ predictors together, the maximum certainty only reaches 88\%.

\begin{figure}[]
\resizebox{\hsize}{!}{\includegraphics[clip=true, width=0.5\textwidth]{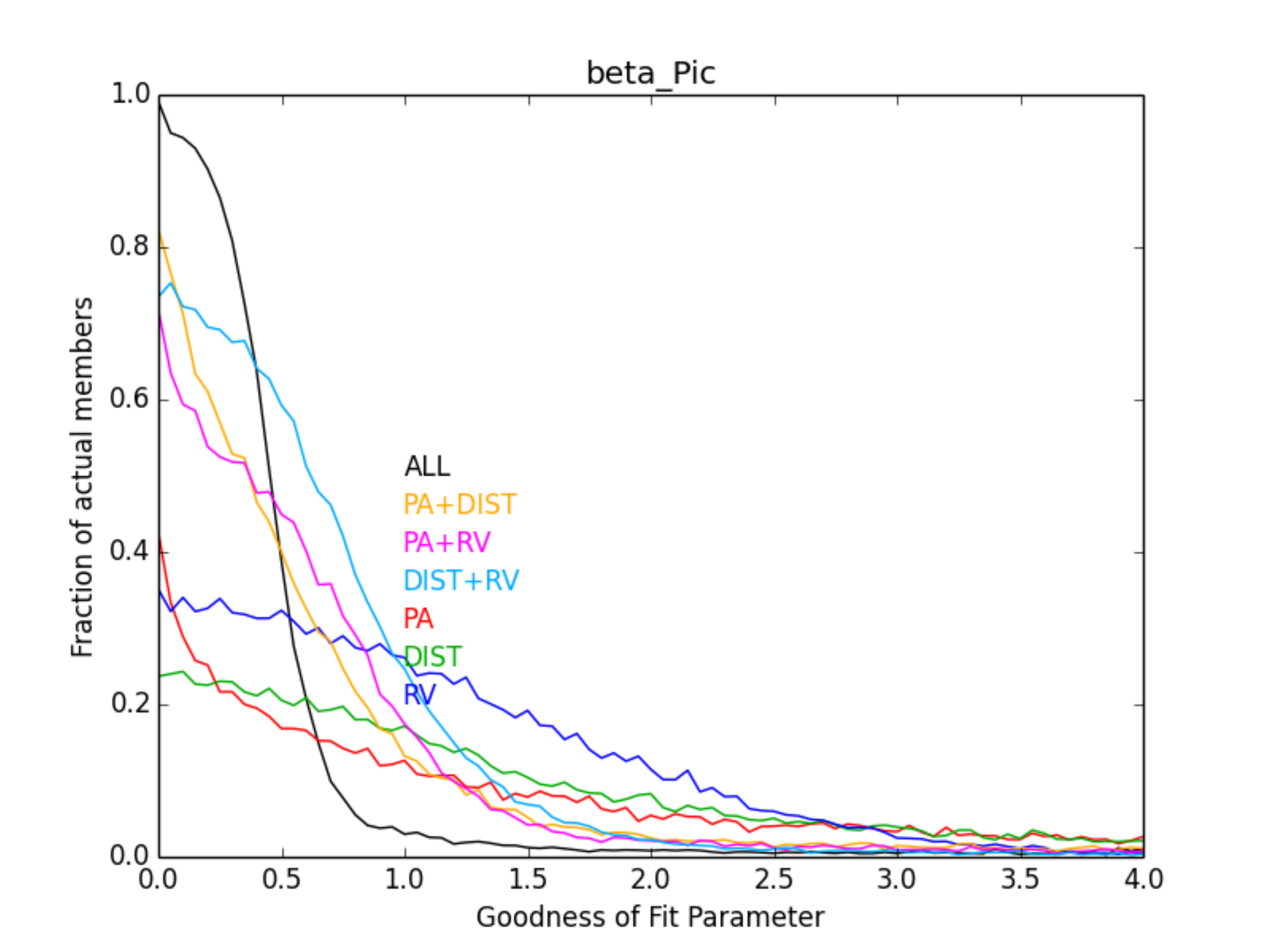}}
\caption{
\footnotesize
Same as Figure \ref{fig:betapic_all} except that f\/ield stars have been removed from consideration, and the probability of an object being a $\beta$ Pictoris member has increased.
}
\label{fig:betapic_young}
\end{figure}

\begin{figure}[]
\resizebox{\hsize}{!}{\includegraphics[clip=true, width=0.5\textwidth]{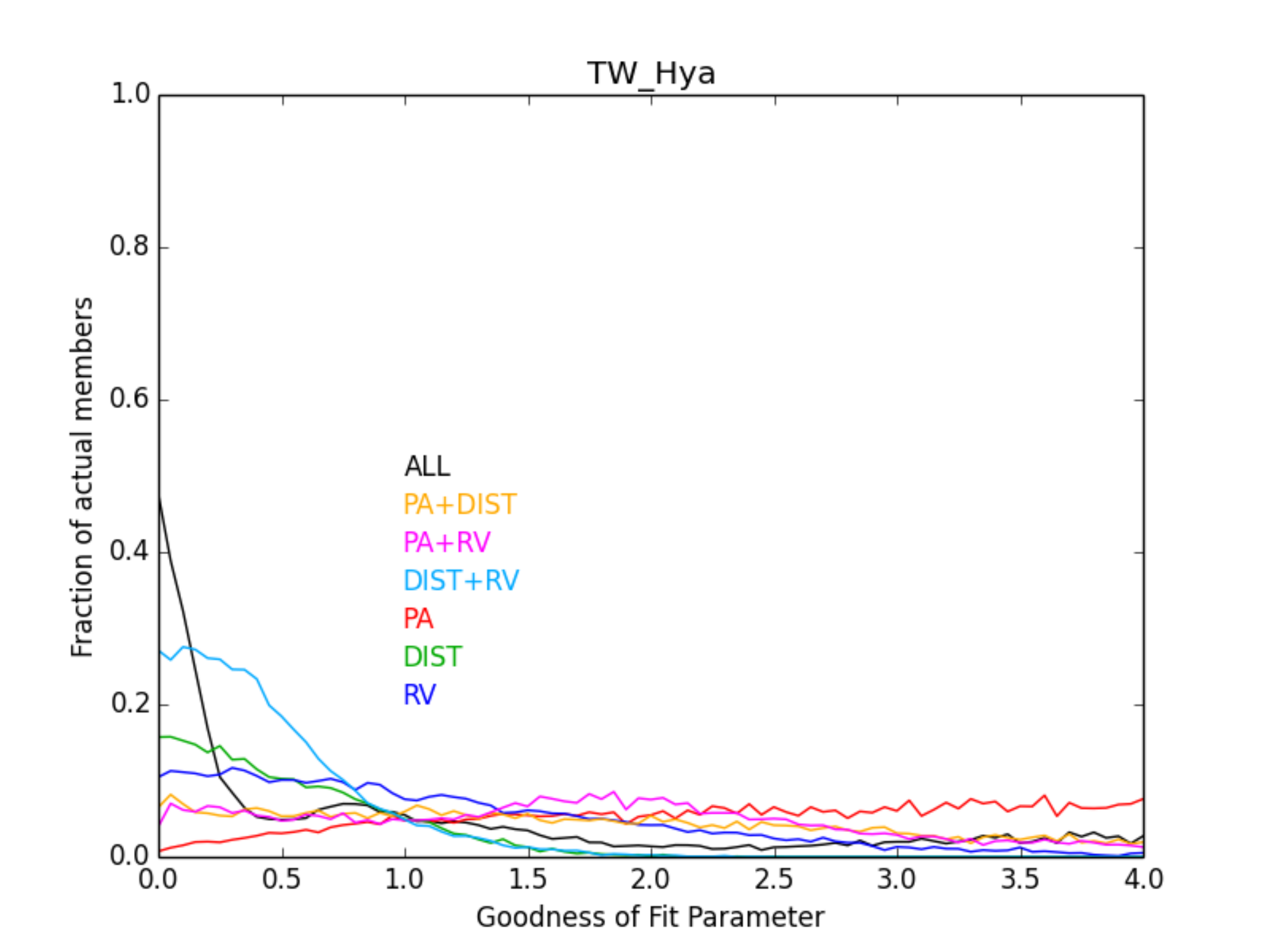}}
\caption{
\footnotesize
Same as Figure \ref{fig:betapic_young} except showing the plot for TW Hydra.  Even without considering f\/ield stars, perfect matches to the TW Hydra moving group are still not likely to be TW Hydra members.
}
\label{fig:twhya_young}
\end{figure}

If f\/ield stars are removed from consideration (a reasonable assumption if it is known that the star is young), the statistics improve signif\/icantly (Figure \ref{fig:betapic_young}).  The result is still not perfect in many cases (Figure \ref{fig:twhya_young}), as several of these distributions overlap, and it is therefore impossible to distinguish between them based on kinematics alone.  In these cases, the code will most often suggest a higher probability of membership in the group that has more members. It is readily apparent from these conclusions that other information is needed, and this kind of kinematics alone is not suff\/icient to determine membership.

\section{Moving Group Tracebacks}
There is one other kinematic technique that can be used to judge the quality of  memberships: kinematic traceback \citep{Makarov2004,Mamajek2013}.  The conceptual underpinning of this technique is that if we assume these groups are the product of a single burst of star formation, they must have been close to each other at the time of formation, where ``close'' is some function of the size of the initial gas cloud.  In practice, this requires all pieces of kinematic information (RA, DEC, $\pi$, $\mu_{RA\cos{DEC}}$, $\mu_{DEC}$, $RV$) for the star and moving group, and some means of approximating galactic orbital motion, whether it's a simple straight-line motion, an epicyclic approximation to galactic motion \citep{Makarov2004}, or a simulation of the galactic gravitational potential into which particles can be placed \citep{Dehnen1998}.  All of these methods make simplif\/ied assumptions about the actual galactic potential f\/ield, neglecting local effects like molecular clouds.

\subsection{Traceback methods}
The particular implementation considered here is an epicyclic approximation of the galactic potential following \citet{Makarov2004}, with updated Oort constants from \citet{Bobylev2010b}.  To calculate the position of the cluster as a function of time, 1000 Monte Carlo points were taken distributed for each of the $N$ bona-f\/ide members of the nearby moving groups.  These points were run back in time from the present to 600 Myr in the past.  At each timestep of 0.1 Myr, freely-oriented ellipsoids were f\/it to each set of $N$ bona-f\/ide members at each timestep.  The mean and standard deviation on the positions and dispersions of the distributions were recorded in a f\/ile.

\begin{figure}[]
\resizebox{\hsize}{!}{\includegraphics[clip=true, width=0.5\textwidth]{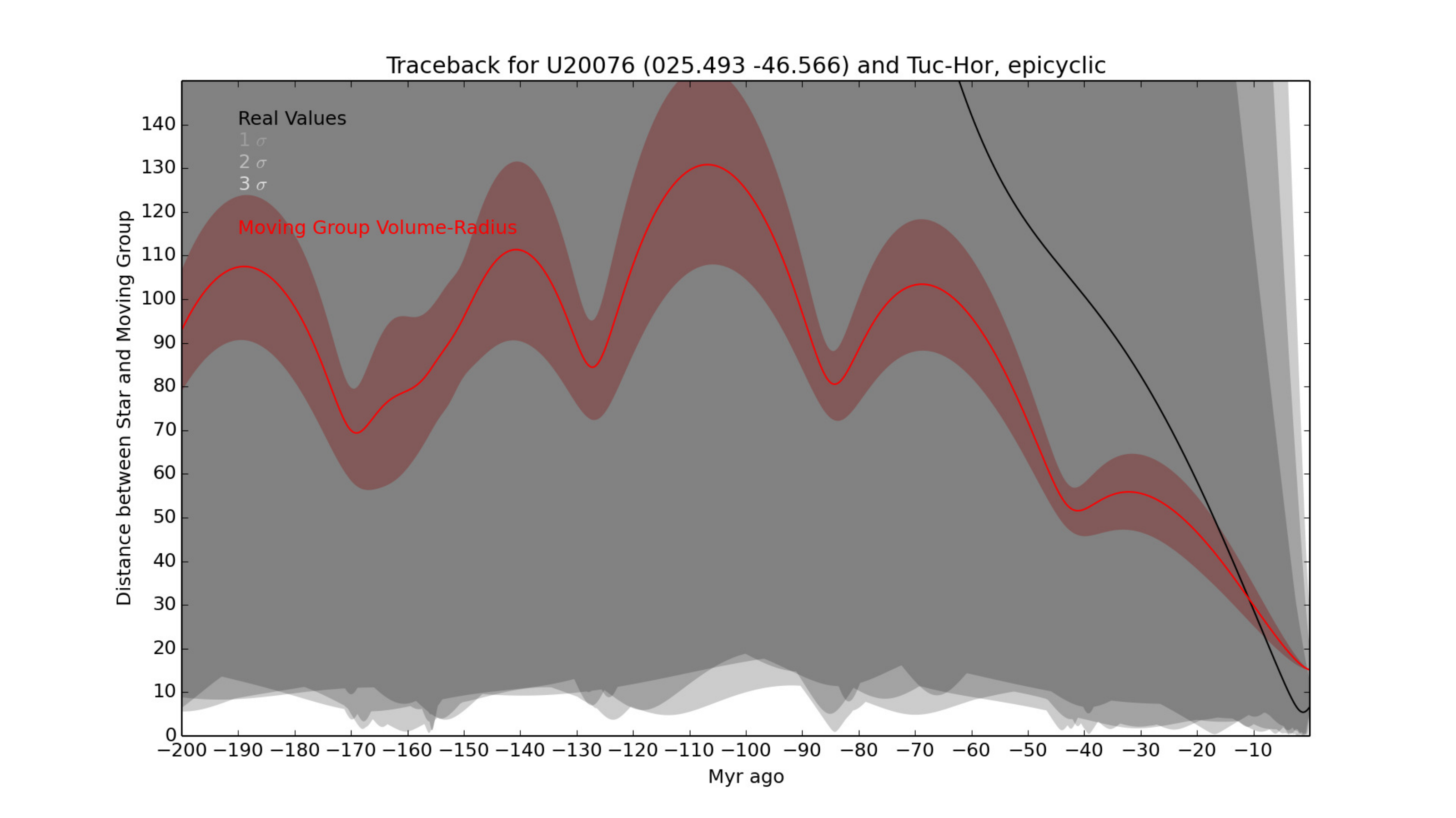}}
\caption{
\footnotesize
A kinematic traceback to the Tucana-Horologium moving group.  Now is on the right-hand side. the position (1$\sigma$, 2$\sigma$, 3$\sigma$) of the brown dwarf is shown in black (dark gray, medium gray, light gray). The effective radius of the moving group (1$\sigma$) is shown in red (pink).  Because the 1$\sigma$ uncertainty envelope is within the pink region 45 Myr ago (the time of formation of Tucana-Horologium), this brown dwarf is potentially a member.
}
\label{fig:traceback_tuchor}
\end{figure}

For each potential young star with full kinematic information, its potential membership was run by computing its own traceback back in time using 20000 Monte carlo points within 1$\sigma$, 2$\sigma$, and 3$\sigma$ of its values.  The separations between the positions of the star and of the moving group are then calculated (Figure \ref{fig:traceback_tuchor}). 

\subsection{Lessons learned from Tracebacks}
\begin{figure}[]
\resizebox{\hsize}{!}{\includegraphics[clip=true, width=0.5\textwidth]{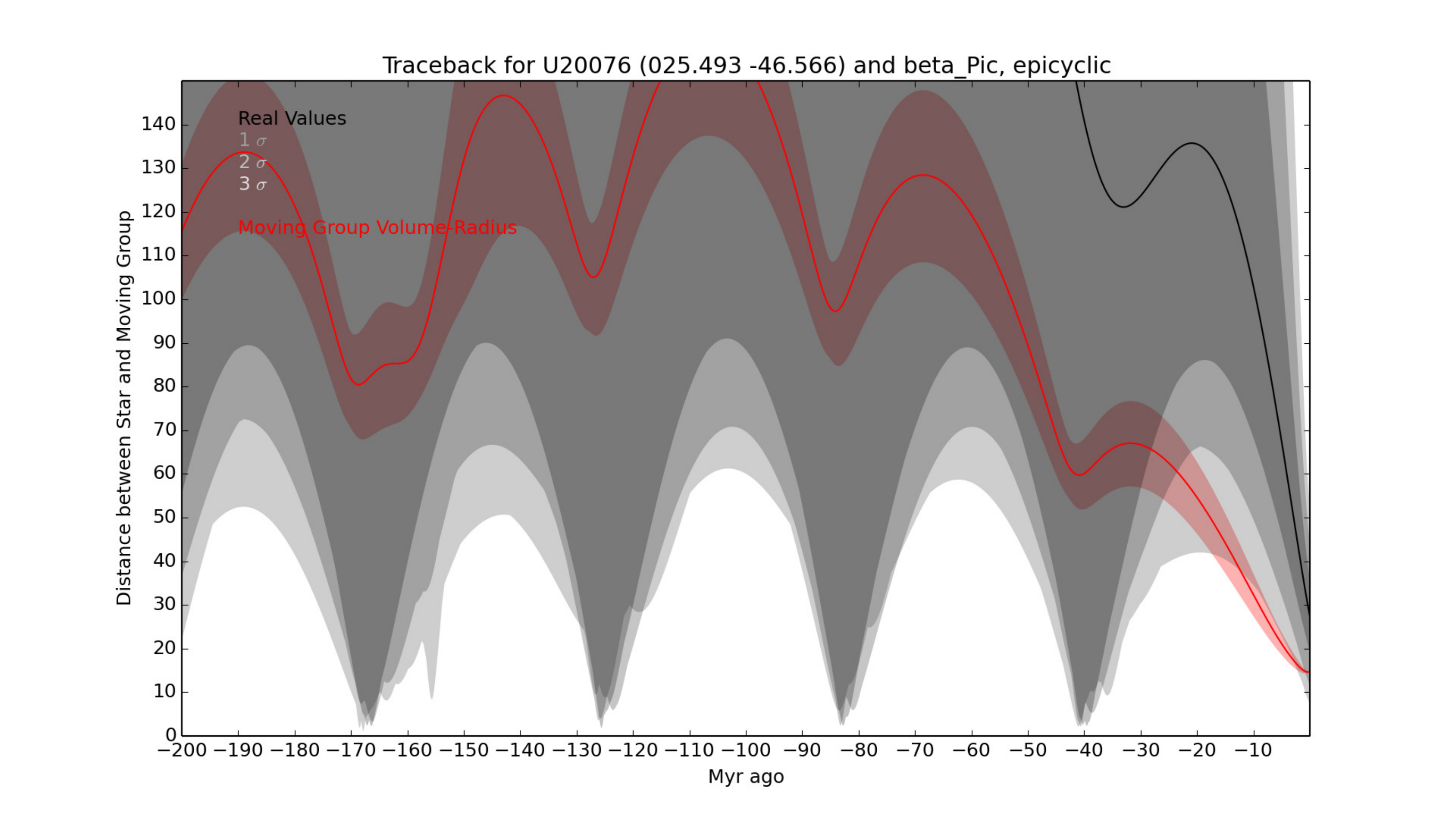}}
\caption{
\footnotesize
A kinematic traceback to the $\beta$ Pictoris moving group, for the same brown dwarf. At the time of formation (25 Myr ago), the brown dwarf position only overlaps with the $\beta$ Pic radius (unrealistically large as it is) at the 2$\sigma$ level; it is therefore not as likely to be a member.
}
\label{fig:traceback_betapic}
\end{figure}

At the moment, large uncertainties on (primarily) stellar parallaxes and radial velocities make kinematic tracebacks less useful than they might otherwise be.  The current state of the art in proper motions, parallaxes, and radial velocities do not constrain the positions of stars very well, with the result that they can appear to be equally good matches to multiple moving groups (see Figure \ref{fig:traceback_betapic}) and therefore have limited discriminative power.

The larger problem inherent in all of this is that the bona-f\/ide members of the young moving groups do not trace back to the same location in space (see F\/Igure 4), leading to enormous apparent sizes at the time of formation, and a generally contracting cluster.  Tucana-Horologium, for example, has an effective radius of over 50 parsecs at t=45 Myr ago, which is enormous compared to the tidal radius of the far more massive Pleiades cluster, 13 pc \citep{Adams2001}, which has presumably not signif\/icantly enlarged since forming from a molecular cloud 125 Myr ago.

\begin{figure}[]
\resizebox{\hsize}{!}{\includegraphics[clip=true, width=0.5\textwidth]{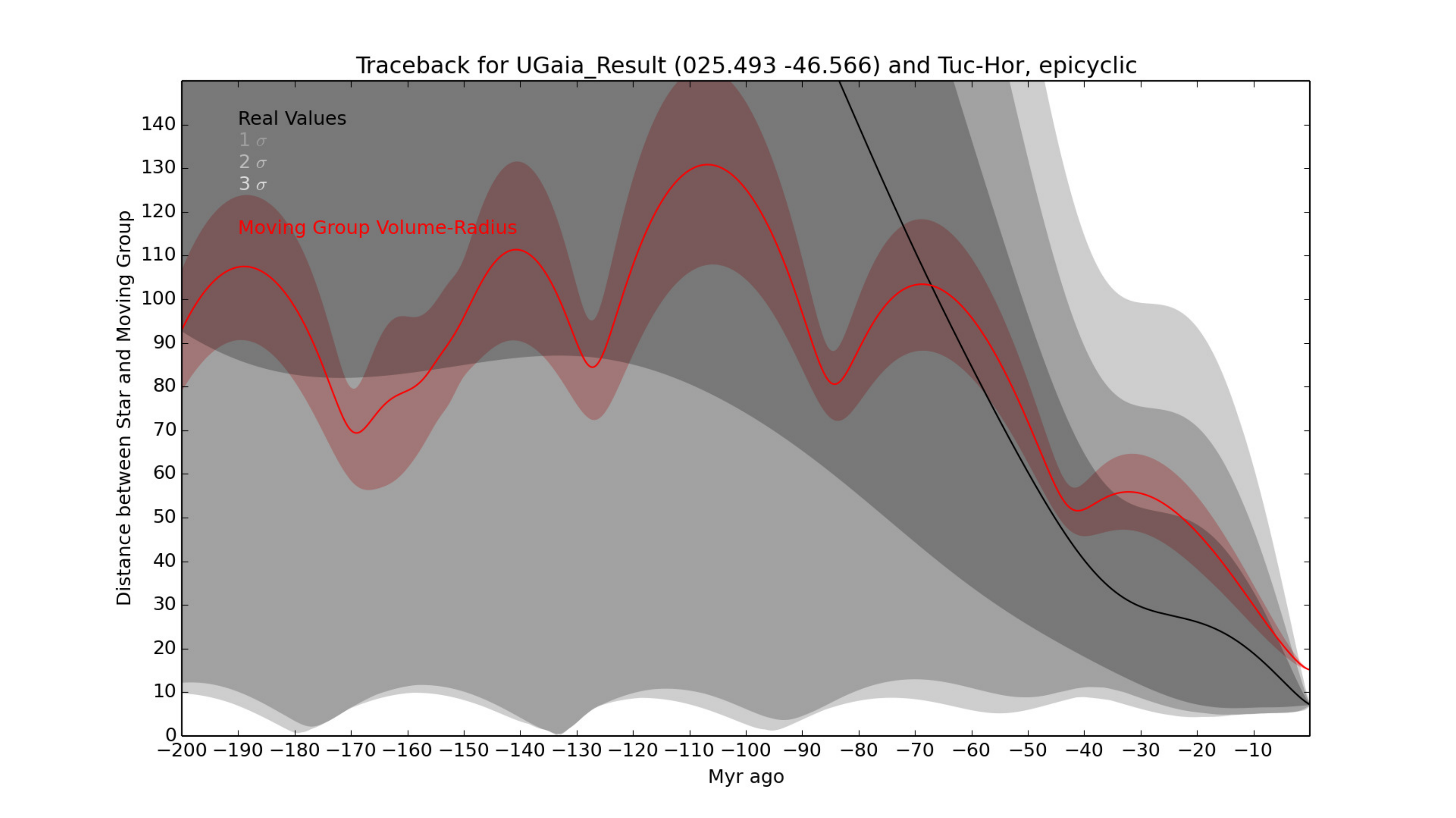}}
\caption{
\footnotesize
The kinematic traceback for the brown dwarf in Figure \ref{fig:traceback_tuchor}, where the astrometry has been replaced by Gaia-quality values that would make it a match, demonstrating the improved constraints possible with Gaia.  The remaining mismatch is entirely due to the radial velocity.
}
\label{fig:traceback_gaia}
\end{figure}

To have a moving group whose members are moving in parallel (as is the case for TW Hydra, \citealt{Weinberger2013}) makes some physical sense, given that the constituent stars are so far apart they have probably never physically interacted.  Some of this discrepancy may be due to the large uncertainties on the data mentioned above, but the more likely conclusion is that some of the stars are physically unrelated, and their current proximity is merely temporary.  Chemical analyses of the members of moving groups (such as AB Dor, \citealt{Barenfeld2013}) show that current member lists are indeed contaminated with non-members, but the youth of the bona-f\/ide members is well-established by other means.  Even if they are not members of a particular group, they are still young stars whose origins need to be explained.  If data of improved precision (Figure \ref{fig:traceback_gaia}) does not resolve the discrepancies, the remaining possibility is that the currently known groups are not the physical entities we thought they were.

\section{The Impact of Gaia}
Many of the moving groups proposed over the years have not stood up to scrutiny. A large part of this problem is that the def\/inition of a moving group has become increasingly strict, going from a mere group of stars with similar space motions, to the group as the product of a single burst of star formation.  A good example of this is the IC 2391 Supercluster from \citet{Eggen1991}, which was identif\/ied on the basis of similar proper motions to the IC 2391 Open Cluster.  Within the paper itself, it is noted that there are multiple subgroups with different ages, but at the time this was not seen as a problem.

The other major change that has happened in the last 20 years is the general availability of milliarcsecond astrometry, largely due to the Hipparcos catalog.  The increased precision makes it possible to identify smaller structures in velocity space, which are more likely to be coeval groups.  \citet{Asiain1999} was one of the f\/irst to apply the new Hipparcos results to the ``Local Association'' and likely pre-discovered the AB Doradus moving group by locating newly visible overdensities in kinematic space.

Gaia will provide a similarly large leap in quality and quantity of available kinematic information.  Gaia's increased precision will allow us to tease out f\/iner, smaller groups that are more likely to be truly co-eval products of a single burst of star formation, and may even completely sweep away the current landscape of nearby young moving groups.  In this way, Gaia will prove fundamental to our understanding of brown dwarfs for which it will not obtain any data. We need this leap, and Gaia's new datasets, to provide a f\/irm foundation for our moving groups so that we can conf\/idently derive brown dwarf properties.


\begin{thebibliography}{14}
\expandafter\ifx\csname natexlab\endcsname\relax\def\natexlab#1{#1}\fi

\bibitem[{{Adams} {et~al.}(2001){Adams}, {Stauffer}, {Monet}, {Skrutskie}, \&
  {Beichman}}]{Adams2001}
{Adams}, J.~D., {Stauffer}, J.~R., {Monet}, D.~G., {Skrutskie}, M.~F., \&
  {Beichman}, C.~A. 2001, \aj, 121, 2053

\bibitem[{{Asiain} {et~al.}(1999){Asiain}, {Figueras}, \& {Torra}}]{Asiain1999}
{Asiain}, R., {Figueras}, F., \& {Torra}, J. 1999, \aap, 350, 434

\bibitem[{{Barenfeld} {et~al.}(2013){Barenfeld}, {Bubar}, {Mamajek}, \&
  {Young}}]{Barenfeld2013}
{Barenfeld}, S.~A., {Bubar}, E.~J., {Mamajek}, E.~E., \& {Young}, P.~A. 2013,
  \apj, 766, 6

\bibitem[{{Bobylev} \& {Bajkova}(2010)}]{Bobylev2010b}
{Bobylev}, V.~V., \& {Bajkova}, A.~T. 2010, \mnras, 408, 1788

\bibitem[{{Dehnen} \& {Binney}(1998)}]{Dehnen1998}
{Dehnen}, W., \& {Binney}, J. 1998, \mnras, 294, 429

\bibitem[{{Eggen}(1991)}]{Eggen1991}
{Eggen}, O.~J. 1991, \aj, 102, 2028

\bibitem[{{Famaey} {et~al.}(2008){Famaey}, {Siebert}, \&
  {Jorissen}}]{Famaey2008}
{Famaey}, B., {Siebert}, A., \& {Jorissen}, A. 2008, \aap, 483, 453

\bibitem[{{Gagn{\'e}} {et~al.}(2014){Gagn{\'e}}, {Lafreni{\`e}re}, {Doyon},
  {Malo}, \& {Artigau}}]{Gagne2014}
{Gagn{\'e}}, J., {Lafreni{\`e}re}, D., {Doyon}, R., {Malo}, L., \& {Artigau},
  {\'E}. 2014, \apj, 783, 121

\bibitem[{{Johnson} \& {Soderblom}(1987)}]{Johnson1987}
{Johnson}, D.~R.~H., \& {Soderblom}, D.~R. 1987, \aj, 93, 864

\bibitem[{{Makarov} {et~al.}(2004){Makarov}, {Olling}, \&
  {Teuben}}]{Makarov2004}
{Makarov}, V.~V., {Olling}, R.~P., \& {Teuben}, P.~J. 2004, \mnras, 352, 1199

\bibitem[{{Malo} {et~al.}(2013){Malo}, {Doyon}, {Lafreni{\`e}re}, {Artigau},
  {Gagn{\'e}}, {Baron}, \& {Riedel}}]{Malo2013}
{Malo}, L., {Doyon}, R., {Lafreni{\`e}re}, D., {Artigau}, {\'E}., {Gagn{\'e}},
  J., {Baron}, F., \& {Riedel}, A. 2013, \apj, 762, 88

\bibitem[{{Mamajek} {et~al.}(2013){Mamajek}, {Bartlett}, {Seifahrt}, {Henry},
  {Dieterich}, {Lurie}, {Kenworthy}, {Jao}, {Riedel}, {Subasavage}, {Winters},
  {Finch}, {Ianna}, \& {Bean}}]{Mamajek2013}
{Mamajek}, E.~E. {et~al.} 2013, \aj, 146, 154

\bibitem[{{Rodriguez} {et~al.}(2013){Rodriguez}, {Zuckerman}, {Kastner},
  {Bessell}, {Faherty}, \& {Murphy}}]{Rodriguez2013}
{Rodriguez}, D.~R., {Zuckerman}, B., {Kastner}, J.~H., {Bessell}, M.~S.,
  {Faherty}, J.~K., \& {Murphy}, S.~J. 2013, \apj, 774, 101

\bibitem[{{Weinberger} {et~al.}(2013){Weinberger}, {Anglada-Escud{\'e}}, \&
  {Boss}}]{Weinberger2013}
{Weinberger}, A.~J., {Anglada-Escud{\'e}}, G., \& {Boss}, A.~P. 2013, \apj,
  762, 118

\end{thebibliography}
\end{document}